\documentclass[journal]{IEEEtran}
\usepackage{enumitem}

\usepackage{amsmath,amssymb,amsfonts,amsthm}
\usepackage{mathtools}

\usepackage{graphicx}
\usepackage{array}
\usepackage{subfigure}
\usepackage{tabularx,booktabs}
\usepackage{siunitx}
\usepackage{multirow}
\usepackage{float}
\usepackage{caption}

\usepackage[table]{xcolor}
\usepackage{soul}

\sethlcolor{yellow!40}               
\soulregister\cite{1}       
\soulregister\url{1}        
\soulregister\ref{1}        
\soulregister\label{1}      
\soulregister\medskip{0}    
\soulregister\noindent{0}   
\soulregister\textbf{1}     
\soulregister\tanh{0}       
\soulregister\mathrm{1}     
\soulregister\bigotimes{0}  
\soulregister\begin{1}      
\soulregister\end{1}        

\usepackage[linesnumbered,ruled]{algorithm2e}
\usepackage{algpseudocode}

\usepackage{cite}
\usepackage{url}

\usepackage{textcomp}
\usepackage{stfloats}
\usepackage{verbatim}
\usepackage{pifont}
\usepackage{braket}
\usepackage{hyperref}
\usepackage{orcidlink}
\usepackage{balance}

\newtheorem{theorem}{Theorem}
\newtheorem{lemma}{Lemma}

\renewcommand{\qedsymbol}{} 

\def\BibTeX{{\rm B\kern-.05em{\sc i\kern-.025em b}\kern-.08em
    T\kern-.1667em\lower.7ex\hbox{E}\kern-.125emX}}

\begin{document}
\title{QSCL-EWIL: Quantum Stochastic Contrast Learning for Enhanced WiFi-Based Indoor Localization
}
\author{Muhammad Bilal Akram Dastagir \orcidlink{0000-0003-2990-4604}, Omer Tariq\,\orcidlink{0000-0002-1771-6166}, Dongsoo Han\,\orcidlink{0000-0001-9844-4221}
\\ Saif~Al-Kuwari\,\orcidlink{0000-0002-4402-7710}, Shahid Mumtaz\orcidlink{0000-0001-6364-6149} and Ahmed Farouk\,\orcidlink{0000-0001-8702-7342} 
\thanks{Muhammad Bilal Akram Dastagir is with the Qatar Center for Quantum Computing, College of Science and Engineering, Hamad Bin Khalifa University, Doha, Qatar.\\
E-mail: mdastagir@hbku.edu.qa}
\thanks{Omer Tariq is with the School of Computing, Korea Advanced Institute of Science and Technology, Daejeon, South Korea.
E-mail: omertariq@kaist.ac.kr}
\thanks{Dongsoo Han is with the School of Computing, Korea Advanced Institute of Science and Technology, Daejeon, South Korea.
E-mail: dshan@kaist.ac.kr}
\thanks{Saif Al-Kuwari is with the Qatar Center for Quantum Computing, College of Science and Engineering, Hamad Bin Khalifa University, Doha, Qatar.
E-mail: smalkuwari@hbku.edu.qa}
\thanks{Shahid Mumtaz is with Nottingham Trent University, Engineering Department, United Kingdom. E-mail: dr.shahid.mumtaz@ieee.org}
\thanks{Ahmed Farouk is with Qatar Center for Quantum Computing, College of Science and Engineering, Hamad Bin Khalifa University, Doha, Qatar, and the Department of Computer Science, Faculty of Computers and Artificial Intelligence, Hurghada University, Hurghada, Egypt.
E-mail: ahmedfarouk@ieee.org}}

\markboth{Journal of \LaTeX\ Class Files,~Vol.~18, No.~9, September~2020}%
{How to Use the IEEEtran \LaTeX \ Templates}

\maketitle

\begin{abstract}
WiFi-based indoor localization is essential for asset tracking, healthcare monitoring, and smart buildings. However, existing systems face challenges such as data variability, environmental noise, and difficulty detecting floor and building levels, compounded by limited labeled data and high received signal strength (RSS) collection costs. This paper introduces quantum stochastic contrast learning (QSCL), a novel framework grounded in rigorous theoretical foundations. We present four theorems and one lemma that establish the probabilistic augmentation, diversity enhancement, relationship preservation, and resilience of QSCL under quantum noise, supported by formal proofs. Leveraging these foundations, QSCL utilizes quantum computing (QC) to generate strong data augmentations with stochastic perturbations, enhancing data diversity, while classical weak augmentations provide subtle variations for robust feature learning. We propose a spatial temporal adaptive attention (STAA) encoder that integrates convolutional layers with adaptive attention mechanisms to capture spatial and temporal dependencies in sequential data. Furthermore, a bidirectional contrastive loss function is introduced to capture forward and reverse relationships between augmented views, ensuring robust representations. Comprehensive evaluations on the UJIIndoorLoc and UTSIndoorLoc datasets validate QSCL, demonstrating superior performance with reduced labeled data and resilience to quantum noise such as bit-flip, dephasing, and measurement noise. The proposed framework significantly improves localization accuracy, floor and building detection, and generalizability in challenging indoor environments.
\end{abstract}

\begin{IEEEkeywords}
WiFi Indoor Localization, Quantum Stochastic Learning, Contrastive Learning Representation, Quantum Augmentation, Flood and Building Detection, Received Signal Strength (RSS).
\end{IEEEkeywords}

\section{Introduction}
\IEEEPARstart{I}{ndoor} localization has emerged as a critical area of research and development due to the exponential growth of mobile devices and the increasing demand for location-based services (LBS). An accurate indoor location is essential for applications such as healthcare care, asset tracking, smart factories, and emergency response \cite{Sridharan2020, BaselineGNNWiFi_Wang2024}. Unlike outdoor systems that rely on GPS, indoor environments pose unique challenges, including signal attenuation and multipath interference, which render GPS ineffective \cite{Luo2019}.

To address these challenges, researchers have increasingly adopted WiFi-based fingerprinting as a cost-effective solution that leverages existing infrastructure \cite{Xu2012}. This approach constructs unique location fingerprints using received signal strength (RSS) from access points (APs), enabling position estimation during the online phase \cite{Kumar2016}. However, RSS measurements are susceptible to environmental factors, such as human movement, structural changes, and signal interference, leading to significant variability that undermines localization accuracy \cite{Li2021}. Thus, addressing noise and variability is critical for developing reliable WiFi-based indoor localization systems \cite{Chen2023}.
Machine learning (ML) and deep learning (DL) methods, including convolutional neural networks (CNNs) and recurrent neural networks (RNNs), have been employed to model RSS patterns and enhance localization performance \cite{He2023}. More recently, graph neural networks (GNNs) have been utilized to capture spatial relationships inherent in RSS data, offering improved robustness in dynamic and noisy environments \cite{Chen2023}. Despite these advancements, existing methods still face challenges in generalizability and handling data variability, particularly in noisy and unseen conditions \cite{He2023}.

Supervised learning approaches are often limited by their reliance on labeled datasets, which are scarce in real-world scenarios. Contrastive learning, by utilizing unlabeled data to learn robust feature representations, has emerged as a promising alternative. By identifying similarities and differences within data, contrastive learning enhances generalization under limited supervision \cite{Chen2020, Khosla2020}. However, current contrastive learning techniques struggle with effective augmentation strategies and fail to model the complex spatio-temporal dependencies needed to address the stochastic nature of noisy environments.

In parallel, the emergence of QC offers transformative potential to tackle these challenges. Quantum properties, such as superposition and entanglement, enable efficient computations that surpass the capabilities of classical approaches \cite{Preskill2018, Arute2019}. In ML, quantum algorithms can accelerate training, optimize feature extraction, and model complex data relationships more effectively \cite{Biamonte2017}. Integrating quantum-assisted learning into classical frameworks enhances generalization \cite{QCHybrid2024} and provides a robust foundation for effective augmentation and representation learning in noisy environments \cite{Gao2022}.

To address the challenges of data variability, environmental noise, and limited labeled data in WiFi-based indoor localization, advancements in augmentation strategies and spatio-temporal encoding are essential. Stochastic augmentations introduce the diversity required for robust learning under variable conditions, while enhanced spatio-temporal models effectively capture dynamic relationships within sequential data. Additionally, a well-designed contrastive loss function ensures meaningful and generalizable feature representations.
This paper introduces Quantum Stochastic Contrastive Learning (QSCL), a novel framework combining rigorous theoretical foundations with empirical validation to advance WiFi-based indoor localization. QSCL is grounded in four theorems and one lemma, supported by formal proofs, which establish the probabilistic diversity of quantum augmentations, resilience to quantum noise, and preservation of meaningful relationships in augmented data. The framework employs quantum-driven strong augmentations alongside classical weak augmentations to enhance robustness and generalization. A Spatio-Temporal Adaptive Attention (STAA) encoder captures spatial and temporal dependencies, while a bidirectional contrastive loss ensures effective representation learning. Empirical evaluations on benchmark datasets, including UJIIndoorLoc \cite{UJIIndoorLoc2014} and UTSIndoorLoc \cite{UTSIndoorLoc2023}, validate QSCL’s superior performance in localization accuracy, distance error estimation, and robustness to noise. By addressing real-world challenges such as data variability and limited labeled data, QSCL provides a scalable and effective solution for practical WiFi-based indoor localization applications, including floor and building detection.

The structure of the paper is as : Section II reviews related work, and Section III presents a detailed methodology of the framework. In Section IV, the impact of quantum noise is presented. Section V provides experimental results and discussions. Finally, Section VI concludes the paper.

\section{Related Work}
The existing research on indoor localization is organized into four main categories: 1) Techniques for fingerprint-based indoor localization, 2) Approaches to reduce RSS uncertainty, 3) Strategies for AP deployment, and 4) Baseline.
\subsection{Fingerprint-Based Indoor Localization Techniques}
For many years, fingerprint-based techniques have been the basis of indoor localization research.
ML algorithms, such as sequential minimum optimization \cite{Varshavsky2007}, decision trees \cite{Bozkurt2015}, and k-nearest neighbors (KNN) \cite{Apostolo2019} have been proposed to enhance the accuracy of RSS fingerprinting. 
Recently, DL has emerged as a promising approach for advancing indoor positioning due to its exceptional feature representation capabilities. Kim et al. \cite{Kim2018} employed stacked autoencoders and feed-forward deep neural networks (DNNs) for floor classification tasks, achieving notable outcomes in extensive indoor areas. Song et al. \cite{Song2019} developed CNNLoc, a one-dimensional CNN that enhanced floor classification and regression performance. Similarly, Cha and Lim \cite{Cha2022} introduced a hierarchical auxiliary DNN (HADNN), which consists of consecutive feed-forward neural networks (CFNNs), to manage complex indoor conditions efficiently. Furthermore, RNNs have been adopted to capture temporal features for scenarios with dynamic user movement. Hoang et al. \cite{Hoang2019} applied long short-term memory (LSTM) and bidirectional LSTM models to improve trajectory tracking and localization precision.
\subsection{Solutions to Mitigate RSS Uncertainty}
RSS-based localization is prone to substantial uncertainty owing to environmental influences such as multipath propagation and signal attenuation. Initial approaches, as the signal propagation models applied by Xu et al. \cite{Xu2012}, are required to forecast the spatial distribution of signals for radio map updates. Nevertheless, these models frequently encountered difficulties in handling complex indoor settings.
To overcome these challenges, He et al. \cite{He2016} introduced a crowdsourcing method where devices functioned simultaneously as “surveyors” and “queriers” to refresh the radio map constantly. Kumar et al. \cite{Kumar2016} applied Gaussian process regression to fill in RSS values, decreasing the manual work necessary to keep the radio map current. Zheng et al. \cite{Zheng2021} utilized heterogeneous GNNs to grasp the spatial closeness between APs and reference points (RPs), thus mitigating RSS fluctuation. In addition, He et al. \cite{He2023} used geomagnetic signals as graph nodes and RSS as connections to form adaptive graph models, using graph attention networks for joint indoor localization.
\subsection{AP Deployment Strategies}
The strategic placement of APs enhances localization precision. Conventional approaches typically aim to extend signal coverage \cite{Du2017}—for example,  Liao et al. \cite{Liao2011} utilized signal coverage thresholds to fine-tune AP placement. In contrast, Du and Yang \cite{Du2017} implemented particle swarm optimization algorithms to optimize the alignment of fingerprints at reference points, facilitating efficient AP positioning.
Luo et al. \cite{Luo2019} investigated the variability of RSS data to determine the best APs. However, methods prioritizing signal reach and fingerprint likeness do not always result in precise localization \cite{Du2017, Zirazi2012}. Zirazi et al. \cite{Zirazi2012} tackled this problem by reducing the geometric dilution of precision, thus improving the accuracy of newly installed APs. Tian et al. \cite{Tian2020} suggested a comprehensive search strategy to find the most effective AP deployment solutions; nevertheless, this method encounters scalability challenges as the number of APs rises.
\subsection{Baseline}
Wang et al. \cite{BaselineGNNWiFi_Wang2024} presented an end-to-end neural network for AP selection designed to optimize AP usage while ensuring high localization precision, thus offering a scalable approach for dynamic indoor settings. Despite its promise, the method faces significant challenges, such as elevated hardware and storage costs due to the large number of APs required and sensitivity to RSS fluctuations from environmental changes. The dependency on RSS for localization presents obstacles in scalability and adaptability to fast-changing environments, affecting real-time accuracy and durability. Moreover, the model's performance may suffer when ground truth data is sparse or nonexistent, limiting its reliability in real-world applications.

\begin{figure*}[t]
    \centering
    \includegraphics[width=\textwidth]{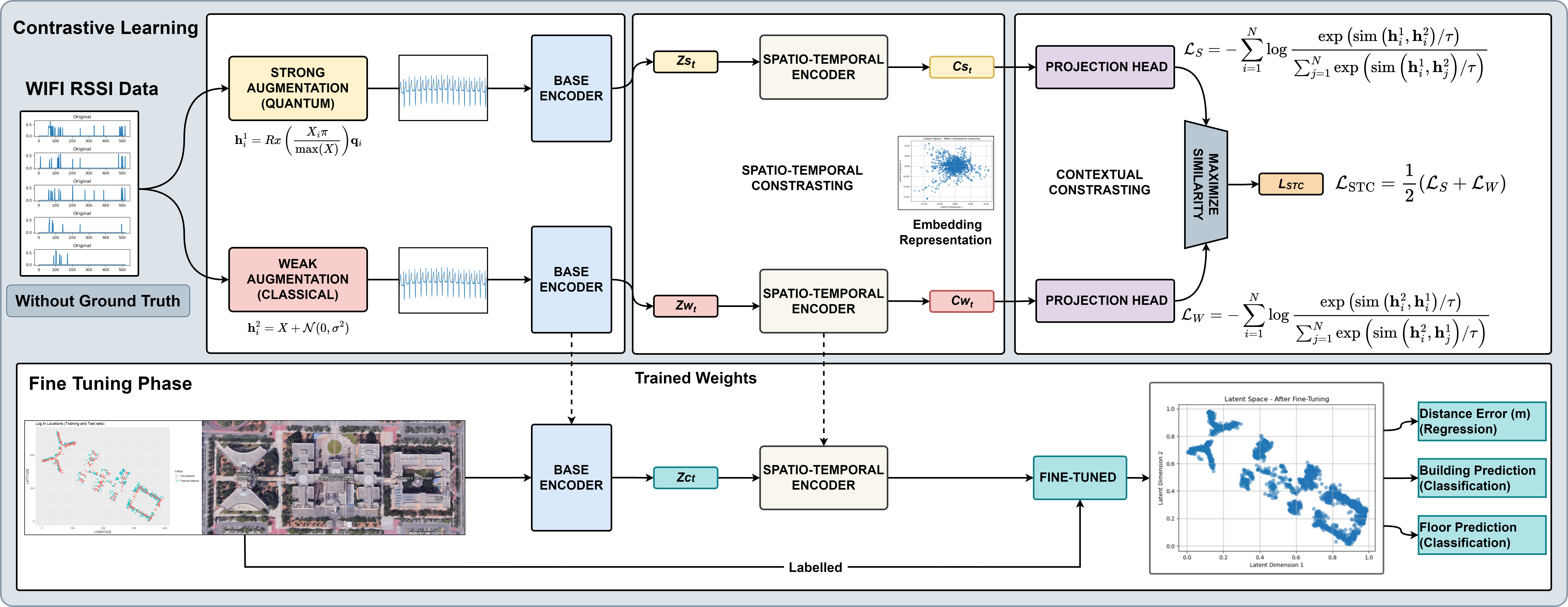} 
    \caption{Proposed quantum stochastic contrastive learning framework for WiFi-based indoor localization.}
    \label{fig:qaa_cl_overview}
\end{figure*}

\section{Methods}
Here, the methodology behind the proposed quantum-assisted augmentation with a contrastive learning framework for WiFi-based indoor localization is discussed in detail. Our approach combines QC techniques with classical ML methods to address the challenges posed by data variability, environmental noise and limited labeled datasets.

\subsection{Problem Formulation and Objective}
WiFi-based indoor location is based on the RSSI values of multiple APs to estimate the position of a device and classify its floor and building within a given structure \cite{Zafari2019, Feng2021WiFiSurvey}. Given unlabeled data for the WiFi signal \( X = \{x_1, x_2, \dots, x_N\} \), where each \(x_i \in \mathbb{R}^d\) represents a feature vector of RSSI values from the APs \(d\), the objective is to learn a mapping ($ f: X \to Y $) which predicts \( Y = (y_{\text{floor}}, y_{\text{building}}, y_{\text{position}}) \) as shown in \textit{Eq.}~\eqref{Eq:Yobj}.

\begin{equation}
Y = (y_{\text{floor}}, y_{\text{building}}, y_{\text{position}}) \in \mathbb{R}^3.
\label{Eq:Yobj}
\end{equation}

The proposed framework employs quantum-assisted augmentation to introduce diversity via strong augmentations using quantum noise operators \( \mathcal{N}_{\text{quantum}}(x_i) \), producing quantum-augmented views \( \mathcal{A}_s(x_i) \) as shown in \textit{Eq.}~\eqref{Eq:QuanAugObj}.

\begin{equation}
\mathcal{A}_s(x_i) = \mathcal{N}_{\text{quantum}}(x_i),
\label{Eq:QuanAugObj}
\end{equation}

To enhance learning, both quantum-assisted strong augmentations \( \mathcal{A}_s \) and classical weak augmentations \( \mathcal{A}_w \) are applied, generating views \( \tilde{x}_i \) and \( \hat{x}_i \) as shown in \textit{Eq.}~\eqref{Eq:augSWObj}.

\begin{equation}
\tilde{x}_i = \mathcal{A}_s(x_i), \quad \hat{x}_i = \mathcal{A}_w(x_i),
\label{Eq:augSWObj}
\end{equation}

The learning objective maximizes the consistency of representations in augmented views using a similarity function \(\text{sim}(\cdot, \cdot)\) and an encoder \( f_{\theta}(\cdot) \), formalized as shown in \textit{Eq.}~\eqref{Eq:augObj}.

\begin{equation}
\max_{\theta} \mathbb{E}\left[\text{sim}(f_{\theta}(\tilde{x}_i), f_{\theta}(\hat{x}_i))\right],
\label{Eq:augObj}
\end{equation}

The contrastive loss function, shown in \textit{Eq.}~\eqref{Eq:contrastiveLossObj}, is designed to minimize intra-cluster similarity while maximizing inter-cluster separation.

\begin{equation}
\min_{\theta} \frac{1}{N} \sum_{i=1}^{N} \log \frac{\exp(\text{sim}(f_{\theta}(x_i), f_{\theta}(x_j)))}{\sum_{k=1}^{N} \exp(\text{sim}(f_{\theta}(x_i), f_{\theta}(x_k)))}.
\label{Eq:contrastiveLossObj}
\end{equation}

Intra-cluster and inter-cluster distances are defined in \textit{Eq.}~\eqref{Eq:intraObj} and \textit{Eq.}~\eqref{Eq:interObj}, respectively.

\begin{equation}
d_{\text{intra}}(x_i, x_j) = \| f(x_i) - f(x_j) \|^2,
\label{Eq:intraObj}
\end{equation}

\begin{equation}
d_{\text{inter}}(x_i, x_k) = \| f(x_i) - f(x_k) \|^2,
\label{Eq:interObj}
\end{equation}

The objective minimizes the ratio of intra-cluster to inter-cluster distance, ensuring robust feature separation, as shown in \textit{Eq.}~\eqref{Eq:minObj}.

\begin{equation}
\text{Minimize} \quad \frac{d_{\text{intra}}}{d_{\text{inter}}}.
\label{Eq:minObj}
\end{equation}

This approach effectively enhances localization accuracy, noise resistance, and generalization in multi-floor and multi-building indoor environments.

\begin{figure*}[t]
    \centering
    \includegraphics[width=\textwidth]{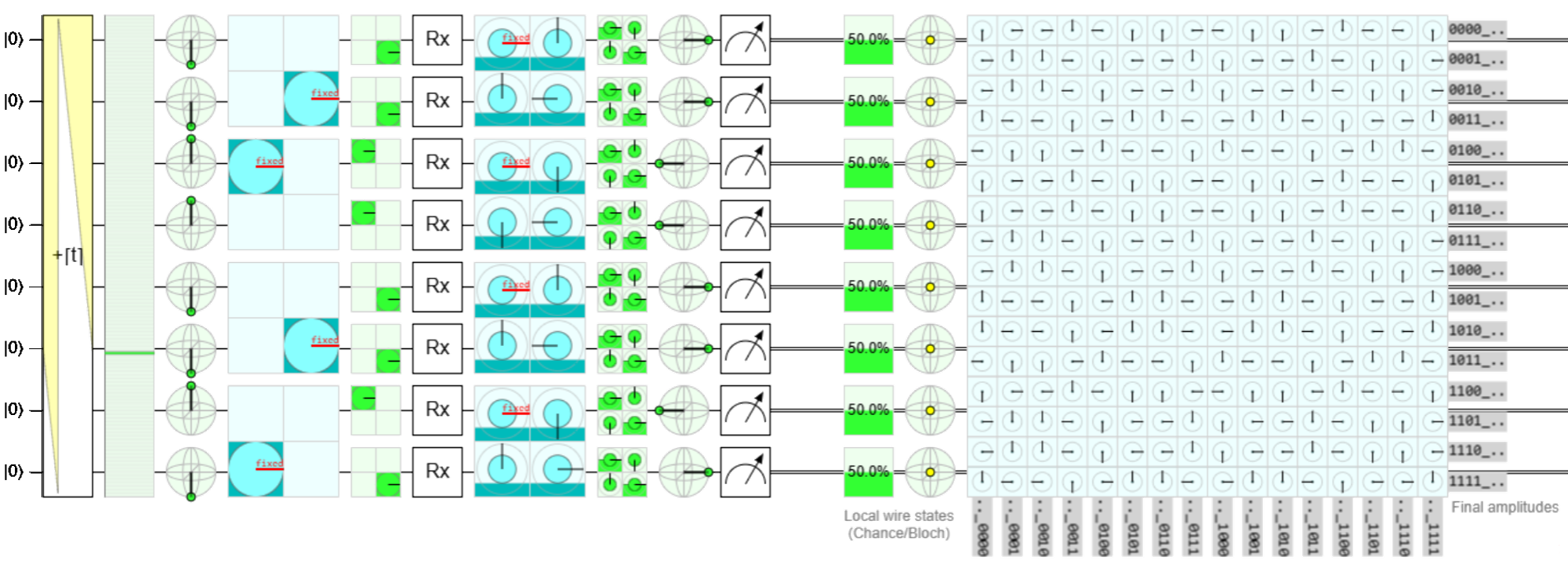} 
    \caption{Proposed quantum-based augmentation circuit.}
    \label{Fig:QAC}
\end{figure*}

\subsection{Strong Quantum Stochasticity Augmentation}
The RSSI (\( X \)), widely used for indoor localization \cite{Zafari2019, Feng2021WiFiSurvey}, is scaled in rotation angles to enable robust and generalizable representations \cite{Biamonte2017, Schuld2015, Chen2020}. The rotation angle \( \theta_i \) for the \( i \)-th RSSI value is calculated as shown in \textit{Eq.} \ref{Eq:theta_i}, where \( X_i \) is the RSSI value for the \( i \)-th  AP and \( \max(X) \) is the maximum RSSI value in the data set. These angles parameterize quantum rotation gates \( R_x(\theta) \), defined as shown in \textit{Eq.} \ref{Eq:rotation_gate}, where \( \sigma_x \) is the Pauli-X operator. We propose a strong quantum-based augmentation as shown in Algorithm \ref{algo:quantum_augmentation}, which is performed using a quantum circuit \( R_{\alpha} \), parameterized by \(\alpha\). The proposed circuit, as shown in Fig. \ref{Fig:QAC}, applies quantum operations such as rotations and entanglement to transform the input RSSI signal \( X \) into an enriched quantum representation. The augmented version \( h_{1}^{S} \) is given and shown in \textit{Eq.} \ref{Eq:lossPropStrong}.
\begin{equation}
    h_{1}^{S} = R_{\alpha}\left(\frac{X_i}{\max(X)}\right) q_i.
    \label{Eq:lossPropStrong}
\end{equation}
where \( q_i \) represents the resulting quantum state after applying the circuit. The normalization term \(\frac{X_i}{\max(X)}\) ensures that the input data is scaled appropriately for quantum processing, maintaining numerical stability and consistency.

\begin{algorithm}[ht!]
\SetAlgoLined
\textbf{Input:} RSSI values $X = \{X_1, X_2, \dots, X_n\}$ from $n$ access points \\
\textbf{Output:} Quantum-augmented view $X_{\text{aug}}$ \\
\textbf{Step 1: Normalize RSSI values}\\
\For{$X_i \in X$}{
    Compute rotation angle: $\theta_i = \frac{X_i \cdot \pi}{\max(X)}$ 
}
\textbf{Step 2: Apply Quantum Circuit}\\
\For{$\theta_i$}{
    Initialize quantum state: $\ket{\psi_i} = \ket{0}$ for qubit $q_i$ \\
    Apply rotation gate: $\ket{\psi_i} \leftarrow R_x(\theta_i) \ket{\psi_i}$ 
}
\textbf{Step 3: Measure Quantum State}\\
\For{$\ket{\psi_i}$}{
    Measure the qubit: $X_{\text{aug}} \gets$ outcome 
}
\Return{$X_{\text{aug}}$}\\
\caption{Quantum-based augmentation of RSSI values.}
\label{algo:quantum_augmentation}
\end{algorithm}

\subsubsection{Theoretical Foundation}
The proposed framework leverages quantum-based augmentation to address challenges in data variability, noise, and generalization for WiFi-based indoor localization. Therefore, we propose the theoretical foundations that establish key theoretical principles. This ensures robustness through probabilistic enhancements as mentioned in Theorem \ref{theorem:probAug}, diversity through quantum noise and measurement randomness as mentioned in Theorem \ref{theorem:divMes}, and semantic consistency by preserving scaled relationships as mentioned in Theorem \ref{theorem:preRel}.

\begin{theorem}
    \textbf{(Probabilistic Augmentation):} Quantum rotation gates probabilistically augment RSSI values with bounded stochastic deviations from the original values.
    \label{theorem:probAug}
    \begin{equation}
        X'_i = X_i \pm \delta, \quad \text{where } |\delta| \leq \frac{\pi}{2}.
        \label{Eq:theorem_probAug}
    \end{equation}
\end{theorem}
\renewcommand{\qedsymbol}{} 
\begin{proof}
 Let \( \theta_i \) be the rotation angle corresponding to a normalized RSSI value \( X_i \), where:
\begin{equation}
    \theta_i = \frac{X_i \cdot \pi}{\max(X)}.
    \label{Eq:theta_i}
\end{equation}
Each RSSI value \( X_i \) is normalized to \( \theta_i \) as in \textit{Eq.}~\eqref{Eq:theta_i} and mapped to the quantum state:
\begin{equation}
    \ket{\psi_i} = R_x(\theta_i) \ket{0},
    \label{Eq:quantum_state}
\end{equation}

The rotation gate \( R_x(\theta_i) \) is given by:
\begin{equation}
    R_x(\theta_i) = e^{-i \frac{\theta_i}{2} \sigma_x} =
    \begin{pmatrix}
        \cos \frac{\theta_i}{2} & -i \sin \frac{\theta_i}{2} \\
        -i \sin \frac{\theta_i}{2} & \cos \frac{\theta_i}{2}
    \end{pmatrix},
    \label{Eq:rotation_gate}
\end{equation}

Applying \( R_x(\theta_i) \) to \( \ket{0} \), the resulting state is:
\begin{equation}
    \ket{\psi_i} = \cos \frac{\theta_i}{2} \ket{0} - i \sin \frac{\theta_i}{2} \ket{1},
    \label{Eq:rotated_state}
\end{equation}

The probabilities of measuring \( \ket{0} \) or \( \ket{1} \) are
\begin{equation}
    P(\ket{0}) = \cos^2 \frac{\theta_i}{2}, \quad P(\ket{1}) = \sin^2 \frac{\theta_i}{2},
    \label{Eq:measurement_outcomes}
\end{equation}

Since \( \cos^2 x + \sin^2 x = 1 \), these probabilities are normalized. During measurement, the qubit collapses to either \( \ket{0} \) with probability 
\( P(\ket{0}) = \cos^2 \frac{\theta_i}{2} \) or \( \ket{1} \) with probability 
\( P(\ket{1}) = \sin^2 \frac{\theta_i}{2} \), depending on the quantum state and the applied rotation angle. The measurement outcome determines the augmented angle \( \theta'_i \), as shown below.
\begin{equation}
    \theta'_i = 
    \begin{cases} 
        2 \arccos \sqrt{P(\ket{0})}, & \text{if the outcome is } \ket{0}, \\
        2 \arcsin \sqrt{P(\ket{1})}, & \text{if the outcome is } \ket{1}.
    \end{cases}
    \label{Eq:augmented_angle}
\end{equation}
Let define the deviation \( \delta \) as shown below.
\begin{equation}
    \delta = \theta'_i - \theta_i,
    \label{Eq:delta}
\end{equation}

From the measurement probabilities, if \( P(\ket{0}) \approx 1 \) (indicating a small rotation), the deviation \( \delta \) approaches zero, resulting in \( \theta'_i \approx \theta_i \), as shown in \textit{Eq.}~\eqref{Eq:small_rotation}.
\begin{equation}
    \delta \approx 0 \quad (\theta'_i \approx \theta_i),
    \label{Eq:small_rotation}
\end{equation}
Conversely, if \( P(\ket{1}) \approx 1 \) (indicating a large rotation), the deviation \( \delta \) approximates \( \pm \frac{\pi}{2} \), leading to \( \theta'_i \neq \theta_i \), as shown in \textit{Eq.}~\eqref{Eq:large_rotation}.
\begin{equation}
    \delta \approx \pm \frac{\pi}{2} \quad (\theta'_i \neq \theta_i),
    \label{Eq:large_rotation}
\end{equation}

Thus, the perturbation \( \delta \) is bounded and shown below.
\begin{equation}
    |\delta| \leq \frac{\pi}{2}.
    \label{Eq:bounded_perturbation}
\end{equation}

Reconstruct the augmented RSSI value \( X'_i \) from \( \theta'_i \), as shown below.
\begin{equation}
    X'_i = \frac{\theta'_i \cdot \max(X)}{\pi},
    \label{Eq:augmented_rssi}
\end{equation}

Given \( \theta'_i \approx \theta_i \pm \delta \), then
\begin{equation}
    X'_i \approx X_i \pm \frac{\delta \cdot \max(X)}{\pi}.
    \label{Eq:augmented_rssi_relationship}
\end{equation}

The stochasticity introduced by quantum measurement ensures that the augmented RSSI value \( X'_i \) probabilistically deviates from the original value \( X_i \), with the relationship and hence validated the theorem and proved as shown in \textit{Eq.}~\eqref{Eq:final_relationship}.
\begin{equation}
    X'_i = X_i \pm \delta.
    \label{Eq:final_relationship}
\end{equation}
This stochastic perturbation \( \delta \) is critical for contrastive learning, as it generates diverse yet meaningful augmented representations in the feature space.
\end{proof}

\begin{theorem}
    \textbf{(Diversity Through Noise and Quantum Measurement):} Quantum measurement randomness and noise ensure that the augmented values are diverse and distinct from the original data.
    \label{theorem:divMes}
    \begin{equation}
    X'_i \neq X_i. 
    \label{Eq:theorem_divMes}
    \end{equation}
\end{theorem}
\renewcommand{\qedsymbol}{} 
\begin{proof}
 Let the augmented value \( X'_i \) be shown below.
\begin{equation}
    X'_i = \frac{\theta'_i \cdot \max(X)}{\pi} + \eta,
    \label{Eq:augmented_value}
\end{equation}

where \( \theta'_i \) is derived from quantum measurement, given and shown below.
\begin{equation}
    \theta'_i = 2 \arcsin \sqrt{P(\ket{1})} \quad \text{or} \quad \theta'_i = 2 \arccos \sqrt{P(\ket{0})},
    \label{Eq:rotation_angle}
\end{equation}

and \( \eta \sim \mathcal{N}(0, \sigma^2) \) represents a noise term. This ensures that the augmented values account for both quantum measurement variability and stochastic noise. The difference between the original RSSI value \( X_i \) and the augmented value \( X'_i \) is:
\begin{equation}
    \Delta X_i = X'_i - X_i = \frac{(\theta'_i - \theta_i) \cdot \max(X)}{\pi} + \eta.
    \label{Eq:difference}
\end{equation}

If \( \theta'_i = \theta_i \), the change in \( X_i \), denoted as \( \Delta X_i \), is determined solely by the noise term \( \eta \), as shown in \textit{Eq.}~\eqref{Eq:noise_only}.
\begin{equation}
    \Delta X_i = \eta.
    \label{Eq:noise_only}
\end{equation}
Since \( \eta \sim \mathcal{N}(0, \sigma^2) \), the probability of \( \Delta X_i = 0 \) is negligible, ensuring that \( X'_i \neq X_i \) in most cases. In contrast, if \( \theta'_i \neq \theta_i \), the change in \( X_i \) incorporates both the deviation between \( \theta'_i \) and \( \theta_i \) and the noise term \( \eta \), as expressed in \textit{Eq.}~\eqref{Eq:quantum_randomness}.
\begin{equation}
    \Delta X_i = \frac{(\theta'_i - \theta_i) \cdot \max(X)}{\pi} + \eta \neq 0.
    \label{Eq:quantum_randomness}
\end{equation}

The augmented value \( X'_i \) remains distinct from \( X_i \), emphasizing the diversity introduced by quantum noise and measurement randomness. The probabilities \( P(\ket{0}) = \cos^2 \frac{\theta_i}{2} \) and \( P(\ket{1}) = \sin^2 \frac{\theta_i}{2} \) inherently introduce stochasticity to \( \theta'_i \), leading to deviations from \( \theta_i \). Even in the absence of noise (\( \eta = 0 \)), this inherent randomness ensures sufficient diversity. Consequently, the interaction between quantum noise \( \eta \) and measurement randomness robustly guarantees that \( X'_i \neq X_i \) in most scenarios.

\end{proof}   

\begin{theorem}
    \textbf{(Preservation of Relationships):} Augmented RSSI values retain a meaningful scaled relationship with the original values, preserving semantic consistency.
    \label{theorem:preRel}
    \begin{equation}
    X'_i = X_i + \Delta.
    \label{Eq:theorem_preRel}
    \end{equation}   
\end{theorem}
\renewcommand{\qedsymbol}{} 
\begin{proof}
Let, the augmented RSSI value \( X'_i \) is defined as:
\begin{equation}
    X'_i = \frac{\theta'_i \cdot \max(X)}{\pi} + \eta,
\end{equation}
where \( \theta'_i \) is the augmented angle, and \( \eta \) is a noise term. The relationship between \( X'_i \) and the original RSSI value \( X_i \) can be expressed as:
\begin{equation}
    X'_i = X_i + \Delta,
\end{equation}
where:
\begin{equation}
    \Delta = \frac{(\theta'_i - \theta_i) \cdot \max(X)}{\pi} + \eta.
\end{equation}
If \( \eta = 0 \) and \( \theta'_i = \theta_i \), then:
\begin{equation}
    \Delta = 0 \implies X'_i = X_i.
\end{equation}
Conversely, if \( \eta \neq 0 \) or \( \theta'_i \neq \theta_i \), then:
\begin{equation}
    \Delta \neq 0 \implies X'_i \neq X_i.
\end{equation}
The perturbation \( \Delta \) is bounded by \textit{Eq.}~\eqref{Eq:BoundedDelta}.
\begin{equation}
    |\Delta| \leq \frac{\max(X)}{\pi} \cdot |\theta'_i - \theta_i| + |\eta|.
    \label{Eq:BoundedDelta}
\end{equation}
ensuring that while \( X'_i \) deviates from \( X_i \), the relationship remains meaningful and scaled. This establishes the bounded stochastic nature of the augmentation process, preserving the semantic relationship between the original and augmented values.
\end{proof}

\subsection{Weak Classical Augmentation}
Weak augmentation employs a classical augmentation strategy that adds Gaussian noise to the original input data, which results in \(h_{2}^{W}\) and is shown in \textit{Eq.}~\eqref{Eq:lossPropWeak}.
\begin{equation}
h_{2}^{W} = X + \mathcal{N}(0, \sigma^2).
\label{Eq:lossPropWeak}
\end{equation}
where \(\mathcal{N}(0, \sigma^2)\) denotes Gaussian noise with zero mean and variance \(\sigma^2\).

\subsection{Encoder Networks}
\subsubsection{Base Encoder}
The base encoder architecture uses a CNN to extract robust feature representations from input sequences. The input data \(X \in \mathbb{R}^{B \times I}\), where \(B\) denotes the batch size and \(I\) denotes the input dimension, are processed by a series of convolutional layers. The base encoder is defined and shown is \textit{Eq.}~\eqref{Eq:baseEncoderEq}.

\begin{align}
X_{\text{cnn}} &= \text{ReLU}(\text{Conv1D}(X, W_{1})), \\
&\xrightarrow{\text{Conv1D}} \text{ReLU}(\text{Conv1D}(X_{\text{cnn}}, W_{2})), \\
&\xrightarrow{\text{Conv1D}} \text{AdaptiveMaxPool1D}(X_{\text{cnn}}).
\label{Eq:baseEncoderEq}
\end{align}
where \(W_{1}\) and \(W_{2}\) are learnable parameters that represent the weights of the convolutional filters, and \(\text{AdaptiveMaxPool1D}\) adjusts the output size to a fixed sequence length \(L\). The resulting output \(X_{\text{cnn}} \in \mathbb{R}^{B \times F \times L}\), with \(F\) denoting the number of filters. 

\subsubsection{Proposed STAA Encoder}
\begin{figure}
    \centering
    \includegraphics[width=\columnwidth]{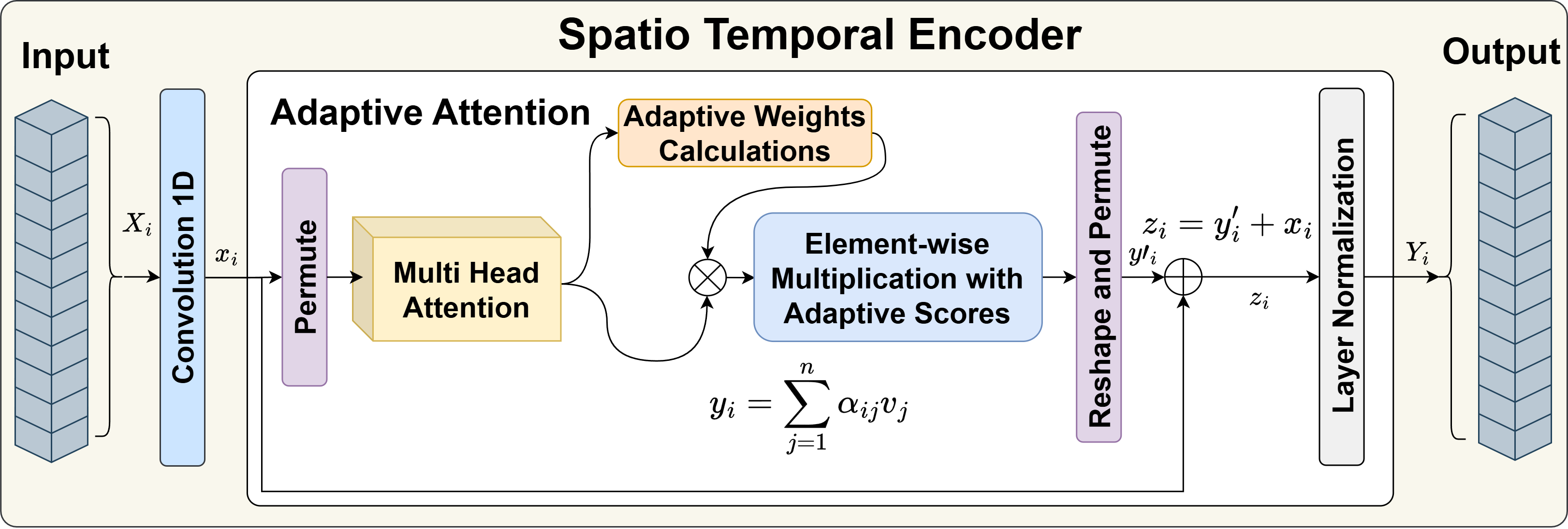} 
    \caption{Proposed spatiotemporal adaptive attention encoder.}
    \label{Fig:STA}
\end{figure}

To capture spatial and temporal relationships in sequential data, we propose the STAA encoder (see Fig. \ref{Fig:STA}). The encoder integrates convolutional operations with adaptive multihead attention mechanisms to effectively learn complex representations. A convolutional layer processes the input sequence \(X_{\text{cnn}}\), followed by a multihead attention mechanism that dynamically assigns importance weights through adaptive scoring. The output of the convolutional layer is shown in \textit{Eq.}~\ref{Eq:STAAEncoderCNN}.

\begin{equation}
X_{\text{conv}} = \text{Conv1D}(X_{\text{cnn}}, W_{\text{conv}}, \text{dilation\_rate}=d),
\label{Eq:STAAEncoderCNN}
\end{equation}
where \(W_{\text{conv}}\) represents the convolutional kernel weights, and \(d\) denotes the dilation rate used for capturing long-range dependencies. Given the output \(X_{\text{conv}}\), a multi-head attention mechanism with adaptive weight scoring is applied. The attention output \(X_{\text{attn}}\) is computed and shown in \textit{Eq.}~\eqref{Eq:STAAEncoderATT}.
\begin{equation}
X_{\text{attn}}, \_ = \text{MHA}(Q = X_{\text{conv}}, K = X_{\text{conv}}, V = X_{\text{conv}}),
\label{Eq:STAAEncoderATT}
\end{equation}
where MHA is multihead attention and an adaptive scoring function \(f_{\text{adapt}}\) computes attention weights as shown in \textit{Eq.}~\eqref{Eq:adaptiveScore}.
\begin{equation}
\alpha_{i} = \sigma(W_{\text{adapt}} \cdot \text{mean}(X_{\text{conv}})),
\label{Eq:adaptiveScore}
\end{equation}
where \(W_{\text{adapt}}\) are learnable weights, \(\sigma\) denotes the sigmoid activation, and \(\alpha_{i}\) are the adaptive attention scores. The final output of the encoder is computed by combining convolutional and attention outputs and shown in \textit{Eq.}~\eqref{Eq:STAAEncoderFinal}.
\begin{equation}
X_{\text{final}} = \text{LayerNorm}(X_{\text{conv}} + \alpha \otimes X_{\text{attn}}).
\label{Eq:STAAEncoderFinal}
\end{equation}

\subsection{Proposed Contrastive Loss Function}

The SimCLR framework utilizes a contrastive loss function to maximize the agreement between different augmentations of the same instance while minimizing the agreement between augmentations of different instances \cite{Chen2020}. The loss is defined and shown in \textit{Eq.}~\eqref{Eq:lossSimCLR}.
\begin{equation}
\mathcal{L}_{i,j} = -\log \frac{\exp(\text{sim}(z_i, z_j)/\tau)}{\sum_{k=1}^{2N} \mathbf{1}_{[k \neq i]} \exp(\text{sim}(z_i, z_k)/\tau)},
\label{Eq:lossSimCLR}
\end{equation}
where \(\text{sim}(z_i, z_j)\) represents the similarity between latent representations \(z_i\) and \(z_j\), \(\tau\) is the temperature parameter, and \(N\) is the batch size.

The SimCLR loss focuses on unidirectional relationships, emphasizing positive pairs while contrasting with negatives, which limits its effectiveness in tasks with complex spatial-temporal dependencies, as only one augmentation contextual is considered while neglecting the second augmentation contextual, therefore lacking diversity and cross-augmentation contextual learning. Therefore, we propose a bidirectional contrastive loss that considers relationships in both directions (\(i \to j\) and \(j \to i\)). Encoder outputs are normalized via a projection head onto a unit hypersphere, ensuring consistency in similarity calculations. The final representations for strong and weak augmentations, \(C_{st}\) and \(C_{wt}\), are optimized with the proposed bidirectional contrastive loss as shown in the \textit{Eq.}~\eqref{Eq:stc_loss}.

\begin{equation}
L_{STC} = \frac{1}{2} (L_S + L_W),
\label{Eq:stc_loss}
\end{equation}
where \(L_S\) and \(L_W\) are the losses for strong and weak augmentations shown in \textit{Eq.}~\eqref{Eq:L_S} and \textit{Eq.}~\eqref{Eq:L_W}.
\begin{equation}
L_S = -\sum_{i=1}^{B} \log \frac{\exp(\text{sim}(C_{st}^i, C_{wt}^i) / \tau)}{\sum_{j=1}^{B} \exp(\text{sim}(C_{st}^i, C_{wt}^j) / \tau)},
\label{Eq:L_S}
\end{equation}
\begin{equation}
L_W = -\sum_{i=1}^{B} \log \frac{\exp(\text{sim}(C_{wt}^i, C_{st}^i) / \tau)}{\sum_{j=1}^{B} \exp(\text{sim}(C_{wt}^i, C_{st}^j) / \tau)}.
\label{Eq:L_W}
\end{equation}

Here, \(B\) is the batch size, \(\text{sim}(\cdot, \cdot)\) denotes a similarity function (e.g., cosine similarity), and \(\tau\) is the temperature parameter. The bidirectional loss ensures consistent and robust embeddings across augmentation.

\begin{algorithm}[ht!]
\SetAlgoLined
\textbf{Input:} Unlabeled WiFi signal data \( X = \{ x_1, x_2, \dots, x_N \} \), labels \( y = \{ y_1, y_2, \dots, y_N \} \) \\
\textbf{Output:} Fine-tuned model \\
\textbf{Hyperparameters:} Temperature \( \tau \), Learning rate \( \eta \), Batch size \( B \), Epochs \( E \) \\
Initialize Base Encoder \( \text{Enc}(\cdot) \), Spatio-Temporal Encoder \( \text{STAA}(\cdot) \), and Projection Head \\
\For{\text{epoch } $e = 1$ to $E$}{
    \For{\text{mini-batch } \( \{x_i\}_{i=1}^B \)}{
        \textbf{Step 1: Generate Augmented Data:}\\ 
        \( h_1^S = R_{\alpha}\left(\frac{x_i}{\max(x)}\right) q_i \) (strong augmentation), \\
        \( h_2^W = x_i + \mathcal{N}(0, \sigma^2) \) (weak augmentation).\\
        \textbf{Step 2: Encode Representations} \\
        \( Z_{st} = \text{Enc}(h_1^S) \), \quad \( Z_{wt} = \text{Enc}(h_2^W) \) \\
        \( C_{st} = \text{STAA}(Z_{st}) \), \quad \( C_{wt} = \text{STAA}(Z_{wt}) \) \\
        \textbf{Step 3: Compute Contrastive Loss} \\
        \( L_S = -\sum_{i=1}^{B} \log \frac{\exp(\text{sim}(C_{st}^i, C_{wt}^i) / \tau)}{\sum_{j=1}^{B} \exp(\text{sim}(C_{st}^i, C_{wt}^j) / \tau)} \) \\
        \( L_W = -\sum_{i=1}^{B} \log \frac{\exp(\text{sim}(C_{wt}^i, C_{st}^i) / \tau)}{\sum_{j=1}^{B} \exp(\text{sim}(C_{wt}^i, C_{st}^j) / \tau)} \) \\
        \( L_{STC} = \frac{1}{2} (L_S + L_W) \) \\
        \textbf{Step 4: Update Weights} \\
        \( \theta \leftarrow \theta - \eta \nabla_\theta L_{STC} \) \\
    }
}
\textbf{Update Weights:} \( \theta \leftarrow \theta - \eta \nabla_\theta L_{\text{STC}} \) \\
\textbf{Fine-Tuning Phase:} $\theta_{\text{pre-trained}} \xrightarrow{\text{weights}} \{\theta_{\text{reg}}, \theta_{\text{class}}\}$\\
\textbf{Regression Loss:} \\\( L_{\text{reg}} = \frac{1}{B} \sum_{i=1}^{B} (y_i - \hat{y}_i)^2 \) \\
\textbf{Classification Loss:} \\\( L_{\text{class}} = -\sum_{i=1}^{B} \sum_{c=1}^{C} y_i^{(c)} \log p_\theta(y_i^{(c)} | x_i) \) 
\caption{Pre-Training Algorithm for Contrastive Learning}
\label{algo:qaa_cl}
\end{algorithm}

\subsection{Proposed Contrastive Learning Representation}
The proposed approach uses two augmentations \(h_1^S\) (strong) and \(h_2^W\) (weak) that are processed through the base encoder and STAA encoder to generate more contextual representations of reverse features and cross-augmentation \(C_{st}\) and \(C_{wt}\) (see Algorithm \ref{algo:qaa_cl}, with a visual representation provided in Fig. \ref{fig:qaa_cl_overview}). These embeddings are normalized onto a hypersphere, facilitating effective similarity calculations in the contrastive loss framework. The contrastive objective ensures robust embeddings across augmented views, improving model performance under noisy conditions. The learned weights are fine-tuned after pretraining for distance regression and floor/building classification. The regression head minimizes the mean squared error using \textit{Eq.}~\eqref{Eq:lossReg}.
\begin{equation}
L_{\text{regression}} = \frac{1}{N} \sum_{i=1}^{N} (y_i - \hat{y}_i)^2,
\label{Eq:lossReg}
\end{equation}
while the classification head minimizes cross-entropy loss is shown in \textit{Eq.}~\eqref{Eq:lossClass}.
\begin{equation}
L_{\text{classification}} = - \sum_{i=1}^{N} \sum_{c=1}^{C} y_i^{(c)} \log p_\theta(y_i^{(c)} | x_i).
\label{Eq:lossClass}
\end{equation}
where \(\lambda_{\text{reg}}\) and \(\lambda_{\text{class}}\) are hyperparameters that balance the contributions of the regression and classification losses, respectively. This formulation enables the model to optimize both tasks effectively, ensuring balanced performance across regression and classification metrics. By integrating quantum-driven augmentations with the STAA encoder, the framework achieves more diverse feature extraction and cross-augmentation contextual learning representations, facilitating accurate and reliable indoor localization even under diverse and noisy conditions.

\section{Impact of Quantum Noise on Quantum Stochasticity Augmentation}
Here, we present novel theoretical foundations to address the challenges posed by quantum noise in NISQ (Noisy Intermediate-Scale Quantum) devices, which inherently affect quantum operations \cite{NISQQC2024}. Our contributions include the introduction of Theorem \ref{theorem:SimConsNoise} and Lemma \ref{lemma:UnbiasedQNoiseCon}, which lay the groundwork for understanding and mitigating the impact of quantum noise on quantum stochasticity augmentation. Quantum noise \cite{ImpactNoiseQC2023}, such as bit-flip, phase-flip, and depolarizing noise, affects the strong augmented view \( C_{st}^i \), while the weak augmented view \( C_{wt}^i \) remains unaffected. The combined effect of these noise types on the quantum state is represented as:
\begin{align}
C_{st}^i &= \mathcal{N}_{\text{bit-flip}}(\ket{\psi_{st}^i}) \nonumber \\
&\quad + \mathcal{N}_{\text{phase-flip}}(\ket{\psi_{st}^i}) \nonumber \\
&\quad + \mathcal{N}_{\text{depolarizing}}(\ket{\psi_{st}^i}),
\label{Eq:CSTNoise}
\end{align}

where noise operators modeled using Pauli matrices, introduce stochastic perturbations. To ensure the robustness of the proposed augmentation framework, Theorem \ref{theorem:SimConsNoise} proves that cosine similarity between the strong augmented view \( C_{st}^i \) and the weak augmented view \( C_{wt}^i \) remains consistent under quantum noise:
\begin{equation}
\text{sim}(C_{st}^i, C_{wt}^i) = \text{sim}((C_{st}^i + \mathcal{N}_{st}), C_{wt}^i).
\label{Eq:TheoremEqToBeProved}
\end{equation}
\textit{Eq.}~\eqref{Eq:TheoremEqToBeProved} guarantees that the augmented views retain their relationships, even when subject to noise, enabling reliable representation learning. Furthermore, Lemma \ref{lemma:UnbiasedQNoiseCon} establishes that quantum noise operators, being stochastic and uncorrelated with the weak augmented view, introduce no systematic bias:
\begin{equation}
\mathbb{E}[\mathcal{N}_{st} \cdot C_{wt}^i] = 0.
\label{Eq:lemma_UnbiasedQNoiseConToBeProved}
\end{equation}
\textit{Eq.}~\eqref{Eq:lemma_UnbiasedQNoiseConToBeProved} ensures that the expected contribution of noise to the similarity computation is null, preserving the integrity of the augmented views.

These theoretical findings, supported by rigorous proofs, form the basis for the proposed quantum stochasticity augmentation framework. They demonstrate how inherent quantum noise in NISQ devices can be effectively managed to ensure robust and consistent representations. Therefore, the proposed framework achieves enhanced performance for WiFi-based indoor localization even under noisy conditions.
\subsection{Theorem and Proof for Impact of Quantum Noise}
\begin{theorem}
    \textbf{(Similarity Consistency Under Quantum Noise):} Cosine similarity between strong and weak augmented views remains consistent even when quantum noise affects the strong augmented view; thus, it satisfies the condition in \textit{Eq.}~\eqref{Eq:theorem_SimConsNoise}.
    \label{theorem:SimConsNoise}
    \begin{equation}
        \text{sim}(C_{st}^i, C_{wt}^i) = \text{sim}((C_{st}^i + \mathcal{N}_{st}), C_{wt}^i).
        \label{Eq:theorem_SimConsNoise}
    \end{equation}
\end{theorem}
\renewcommand{\qedsymbol}{} 
\begin{proof}
    Let \( C_{st}^i \) represent the strong augmented view of a quantum state, where the classical data are transformed into quantum states via quantum operations. Similarly, let \( C_{wt}^i \) represent the weak augmented view, which remains unaffected by noise, then quantum noise \( \mathcal{N}_{st} \), representing bit-flip, phase-flip, or depolarizing noise, is applied to the strong augmented view. The strong and noisy augmented view is defined and shown in \textit{Eq.}~\eqref{Eq:CstNoise}.
\begin{equation}
C_{st}^i = C_{st}^i + \mathcal{N}_{st},
\label{Eq:CstNoise}
\end{equation}
where \( \mathcal{N}_{st} \) is the noise operator applied to the quantum state. The cosine similarity between the strong and weak augmented views is used as the measure of similarity:
\begin{equation}
\text{sim}(A, B) = \frac{A \cdot B}{\| A \| \| B \|},
\label{Eq:cosineSimilarity}
\end{equation}
where \( A \cdot B \) denotes the dot product, and \( \| A \|, \| B \| \) are the Euclidean norms of the vectors \( A \) and \( B \). For augmented views \( C_{st}^i \) and \( C_{wt}^i \), the similarity becomes:
\begin{equation}
\text{sim}(C_{st}^i, C_{wt}^i) = \frac{C_{st}^i \cdot C_{wt}^i}{\| C_{st}^i \| \| C_{wt}^i \|},
\label{Eq:augmentedSimilarity}
\end{equation}

When quantum noise \( \mathcal{N}_{st} \) is applied, the similarity is updated as:
\begin{equation}
\text{sim}((C_{st}^i + \mathcal{N}_{st}), C_{wt}^i) = \frac{(C_{st}^i + \mathcal{N}_{st}) \cdot C_{wt}^i}{\| C_{st}^i + \mathcal{N}_{st} \| \| C_{wt}^i \|},
\label{Eq:noiseSimilarity}
\end{equation}

Expanding the numerator:
\begin{equation}
(C_{st}^i + \mathcal{N}_{st}) \cdot C_{wt}^i = C_{st}^i \cdot C_{wt}^i + \mathcal{N}_{st} \cdot C_{wt}^i,
\label{Eq:numeratorExpansion}
\end{equation}

Given that quantum noise operators such as bit-flip, phase-flip, or depolarizing noise act orthogonally to the original quantum state, the term \( \mathcal{N}_{st} \cdot C_{wt}^i \) approaches zero from \textit{Lemma} \ref{lemma:UnbiasedQNoiseCon}.
\begin{equation}
\mathbb{E}[\mathcal{N}_{st} \cdot C_{wt}^i] = 0,
\label{Eq:noiseExpectation}
\end{equation}

Thus, the similarity simplifies to:
\begin{equation}
\text{sim}((C_{st}^i + \mathcal{N}_{st}), C_{wt}^i) \approx \text{sim}(C_{st}^i, C_{wt}^i).
\label{Eq:finalSimilarity}
\end{equation}

Therefore, Theorem \ref{theorem:SimConsNoise} is proved and can be written as shown in \textit{Eq.}~\eqref{Eq:TheoremEqProved}. 

\begin{equation}
\text{sim}(C_{st}^i, C_{wt}^i) = \text{sim}((C_{st}^i + \mathcal{N}_{st}), C_{wt}^i)
\label{Eq:TheoremEqProved}
\end{equation}

\textit{Eq.}~\eqref{Eq:TheoremEqProved} validated that the similarity remains balanced and consistent, even in quantum noise. This shows the resilience of our model to quantum noise (bit-flip, dephasing, and measurement noise), which overcomes noise-related issues in a noisy environment.
\end{proof}

\begin{lemma}
    \textbf{(Unbiased Neutrality of Quantum Noise Operator):} Quantum noise operators, being stochastic and uncorrelated with weak augmented views, introduce no systematic bias in the similarity computation.
    \label{lemma:UnbiasedQNoiseCon}
    \begin{equation}
        \mathbb{E}[\mathcal{N}_{st} \cdot C_{wt}^i] = 0.
        \label{Eq:lemma_UnbiasedQNoiseCon}
    \end{equation} 
\end{lemma}
\begin{proof}
Let the weak augmented view \( C_{wt}^i \) be a classical vector and remain unaffected by quantum noise. The strong augmented view \( C_{st}^i \) is noisy by a quantum noise operator \( \mathcal{N}_{st} \), which represents the perturbations introduced by quantum processes such as bit-flip, phase-flip, or depolarizing noise. The first term, \( C_{st}^i \cdot C_{wt}^i \), then represents the expected similarity between the strong and uncorrupted augmented view and the weak augmented view. The second term, \( \mathcal{N}_{st} \cdot C_{wt}^i \), quantifies the effect of quantum noise on the similarity computation. To ensure that the similarity computation remains unbiased, the expectation of the second term is analyzed:
\begin{equation}
\mathbb{E}[\mathcal{N}_{st} \cdot C_{wt}^i],
\end{equation}
Assume that the quantum noise \( \mathcal{N}_{st} \) is uncorrelated with the weak augmented view \( C_{wt}^i \) which means that:
\begin{equation}
\mathbb{E}[\mathcal{N}_{st} \cdot C_{wt}^i] = \mathbb{E}[\mathcal{N}_{st}] \cdot C_{wt}^i,
\end{equation}
For typical quantum noise processes, such as bit-flip, phase-flip, or depolarizing noise, the expectation of the noise operator is zero \cite{NoiseAnalysisSebastian2023} due to its stochastic and unbiased nature:
\begin{equation}
\mathbb{E}[\mathcal{N}_{st}] = 0,
\end{equation}
Therefore,
\begin{equation}
\mathbb{E}[\mathcal{N}_{st} \cdot C_{wt}^i] = 0.
\label{Eq:lemmaFinalProof}
\end{equation}

\textit{Eq.}~\eqref{Eq:lemmaFinalProof}  shows that the quantum noise does not introduce systematic bias into the similarity computation, as the expected contribution of \( \mathcal{N}_{st} \cdot C_{wt}^i \) is zero.

\end{proof}

\section{Experimental Validation and Discussion}
The proposed quantum-assisted augmentation framework comprehensively evaluates and highlights its effectiveness in WiFi-based indoor localization tasks. Using quantum-enhanced augmentations, the framework demonstrates significant improvements in localization accuracy. The experimental results are validated on two benchmark datasets, UJIIndoorLoc \cite{UJIIndoorLoc2014} and UTSIndoorLoc \cite{UTSIndoorLoc2023}, using a combination of contrastive learning and fine-tuning for regression and classification tasks with performance evaluated against state-of-the-art methods for indoor localization.

\subsection{WiFi Datasets, Pre-processing and Experimental Setup}
The UJIIndoorLoc dataset contains 2,000 labeled instances collected on a university campus, featuring data from 52 WiFi APs. Each instance includes RSSI values, Floor\_ID, Building\_ID, and position identifiers. The data set is evenly divided into 1,000 training samples and 1,000 testing samples, with the primary objective being to predict the position of the building, floor, and device based on WiFi signals. Preprocessing for UJIIndoorLoc dataset involves mean imputation to handle missing RSSI values, min-max scaling to normalize signal strength within the range [0, 1], and one-hot encoding for categorical features such as Floor\_ID and Building\_ID. Instances with RSSI values below -90 dBm are discarded to reduce noise and enhance data quality. The UTSIndoorLoc dataset, collected within the University of Technology Sydney, comprises 5,000 WiFi RSSI data from 38 WiFi APs across four floors. It includes additional environmental variables, presenting a more complex scenario than UJIIndoorLoc. Similar preprocessing steps are applied: mean imputation addresses missing RSSI values, min-max normalization ensures uniform signal strength ranges, and one-hot encoding is used for Floor\_ID and Building\_ID. To mitigate the impact of unreliable measurements, instances with RSSI values below -90 dBm are removed.

The experiment was conducted on a hardware setup featuring an Intel i9 CPU running at 3.20 GHz, 32GB of DDR5 RAM, and an RTX 4090 GPU. This setup ensures efficient training and evaluation of the DL models, particularly for the high computational demands of contrastive learning and quantum-enhanced augmentation processes. The model implementation uses TensorFlow and qiskit with GPU acceleration for optimized performance during training and inference.



\begin{figure*}[t]
    \centering
    \includegraphics[width=\textwidth]{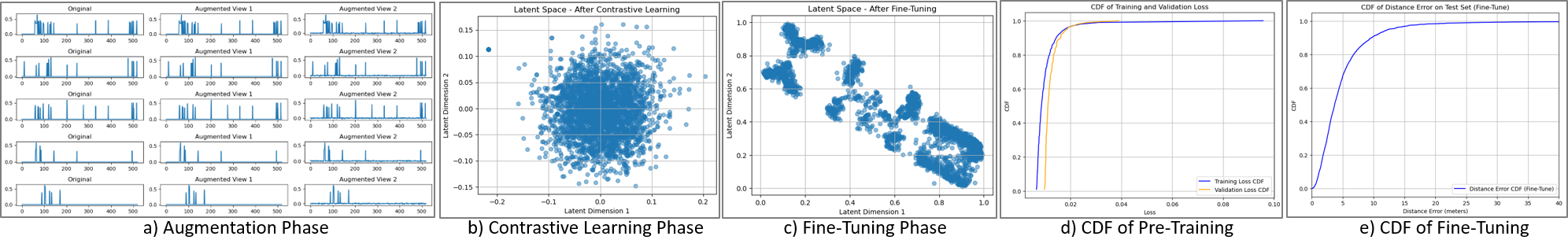}  
    \caption{Propose framework implementation result stepwise.}
    \label{fig:impFrameEvaluationResult}
\end{figure*}
\subsection{Implementation Framework and Evaluation Results}
The proposed method’s implementation framework and evaluation process are shown in Fig. \ref{fig:impFrameEvaluationResult}. It consists of data augmentation (Fig. \ref{fig:impFrameEvaluationResult}-a), training using contrastive learning (Fig. \ref{fig:impFrameEvaluationResult}-b), fine-tuning (Fig. \ref{fig:impFrameEvaluationResult}-c), and performance assessment in terms of CDF (Fig.\ref{fig:impFrameEvaluationResult}-d and Fig. \ref{fig:impFrameEvaluationResult}-e). The input data undergo quantum-assisted augmentation, generating two views, original and augmented (View 1 and View 2). These views are processed through the contrastive learning network, which optimizes the representations by clustering similar views and separating dissimilar ones in the latent space. The learned latent representations are visualized in two dimensions during the training and fine-tuning phases. Initially, the latent space after contrastive learning shows scattered points, indicating the model's ability to map complex input data into a structured lower-dimensional space. Fine-tuning further refines this latent space, resulting in more distinct clusters, essential for accurate multi-floor and building localization. Performance is evaluated using a cumulative distribution function (CDF) curve of the distance error on the test set. The CDF illustrates the probability of achieving specific distance errors, with a steep rise indicating low errors with high probability. The proposed framework demonstrates superior localization accuracy, as evidenced by a sharp increase in the CDF curve, outperforming traditional methods. These results validate the robustness and effectiveness of the model in accurately predicting locations in diverse indoor environments.

\begin{figure*}[t]
    \centering
    \includegraphics[width=\textwidth]{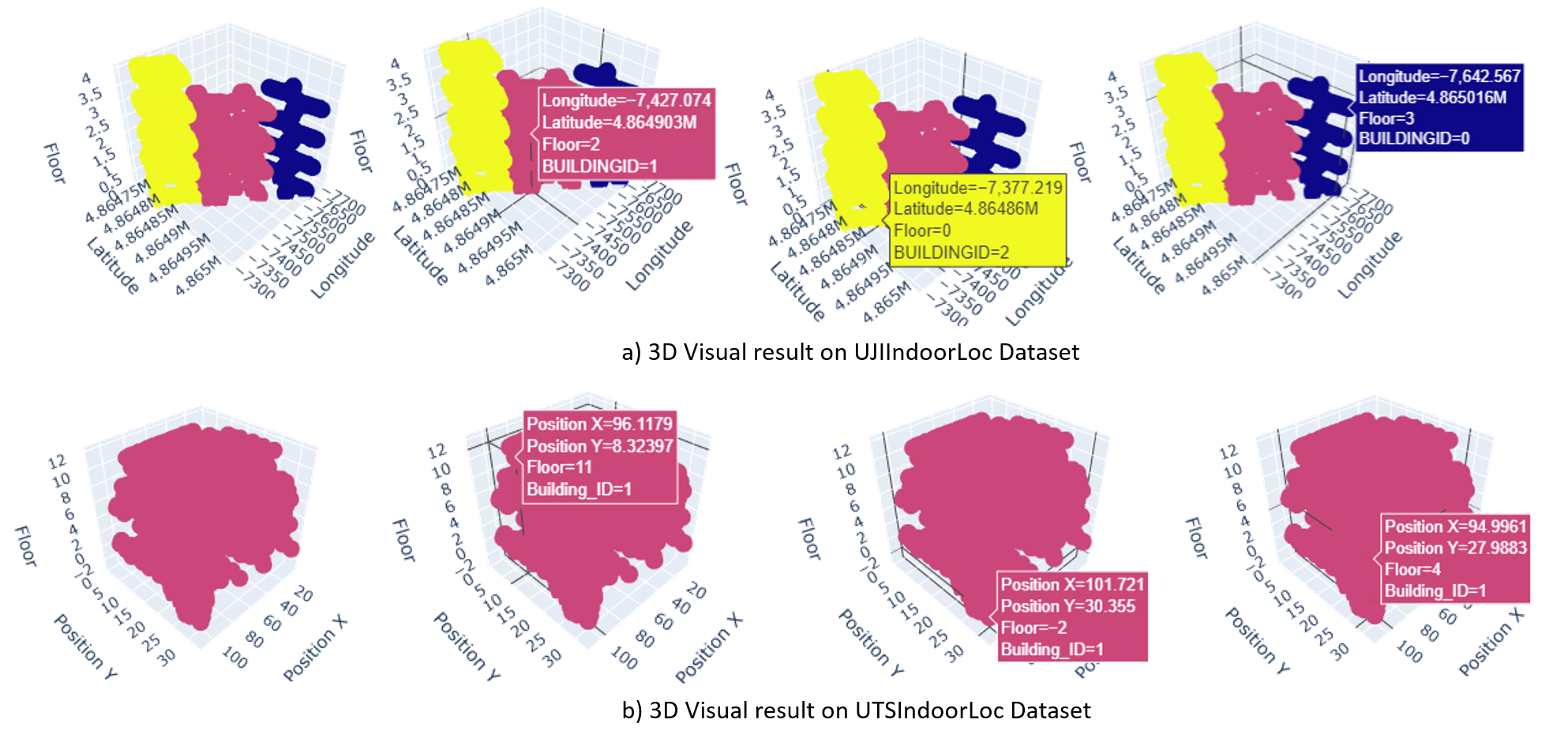} 
    \caption{3D visualization of the resultant accuracy for WiFi-based indoor localization with floor and building detection on a) UJIIndoorLoc dataset and b) UTSIndoorLoc dataset.}
    \label{fig:3DVisualResults}
\end{figure*}

\begin{table*}[ht]
\centering
\begin{tabular}{|c|c|c|c|c|c|}
\hline
\multirow{2}{*}{\textbf{Method}} & \multicolumn{3}{c|}{\textbf{UJIIndoorLoc Dataset}} & \multicolumn{2}{c|}{\textbf{UTSIndoorLoc Dataset}} \\ \cline{2-6} 
 & \textbf{Building Accuracy} & \textbf{Floor Accuracy\%} & \textbf{Mean Location Error (m)} & \textbf{Floor Accuracy \%} & \textbf{Mean Location Error (m)} \\ \hline
\textbf{Random Forest} & 99.95 & 79.92 & 16.73 & 89.17 & 8.71 \\ \hline
\textbf{Decision Tree} & 98.91 & 78.03 & 22.89 & 77.31 & 12.69 \\ \hline
\textbf{KNN\cite{Kim2018}} & 99.82 & 92.16 & 12.66 & 98.91 & 8.58 \\ \hline
\textbf{DNN\cite{Kim2018}} & 99.10 & 79.12 & 17.53 & 97.16 & 9.26\\ \hline
\textbf{CNNLoc\cite{Song2019}} & 99.27 & 96.03 & 11.78 (34.85) & 94.57 & 7.60 (9.77) \\ \hline
\textbf{HADNNh\cite{Cha2022}} & 99.82 & 93.88 & 11.27 & 87.11 & 10.62 \\ \hline
\textbf{HADNN\cite{Cha2022}} & 100 & 93.15 & 11.59 & 92.01 & 10.36 \\ \hline
\textbf{CFNN\cite{Cha2022}} & 100 & 92.34 & 11.74 & 89.17 & 10.32 \\ \hline
\textbf{GNN (baseline)\cite{BaselineGNNWiFi_Wang2024}} & 100 & 94.15 & 9.61 & 97.91 & 7.48 \\ \hline
\textbf{Proposed} & \textbf{99.37 $\pm$ 0.63} & \textbf{96.89 $\pm$ 0.82} & \textbf{6.29 $\pm$ 0.57} & \textbf{98.12 $\pm$ 0.783} & \textbf{4.635 $\pm$ 0.825} \\ \hline
\end{tabular}
\vspace{6pt}
\caption{Comparison of Proposed Approach with State-of-the-Art Works on Two Public Datasets}
\label{table:compSOTA}
\end{table*}

\begin{table*}[htbp]
\centering
\begin{tabular}{|c|c|c|c|c|c|c|}
\hline
\multirow{2}{*}{\textbf{AP Selection Method}} & \multirow{2}{*}{\textbf{Position Method}} & \multicolumn{3}{c|}{\textbf{UJIIndoorLoc Dataset}} & \multicolumn{2}{c|}{\textbf{UTSIndoorLoc Dataset}} \\ \cline{3-7} 
 &  & \parbox[t]{0.7in}{\centering \textbf{Building} \\ \textbf{Accuracy (\%)}} & \parbox[t]{0.7in}{\centering \textbf{Floor} \\ \textbf{Accuracy (\%)}} & \parbox[t]{1in}{\centering \textbf{Mean Location} \\ \textbf{Error (m)}} & \parbox[t]{0.7in}{\centering \textbf{Floor} \\ \textbf{Accuracy (\%)}} & \parbox[t]{1in}{\centering \textbf{Mean Location} \\ \textbf{Error (m)}} \\ \hline
\textbf{AP Selection Network} & \textbf{Our Approach} & \textbf{99.37 $\pm$ 0.63} & \textbf{96.89 $\pm$ 0.82} & \textbf{6.29 $\pm$ 0.57} & \textbf{98.12 $\pm$ 0.78} & \textbf{4.64 $\pm$ 0.83} \\ \hline
 & \textbf{GNN\cite{BaselineGNNWiFi_Wang2024}} & 100 & 92.82 & 9.61 & 97.19 & 7.87 \\ \hline
 & \textbf{CFNN\cite{Cha2022}} & 99.82 & 91.01 & 10.97 & 87.62 & 9.72 \\ \hline
 & \textbf{HADNN\cite{Cha2022}} & 100 & 91.54 & 11.05 & 91.75 & 9.17 \\ \hline
 & \textbf{CNNLoc\cite{Song2019}} & 99.87 & 92.80 & 35.67 & 94.08 & 10.29 \\ \hline
\textbf{AP Threshold Selection} & \textbf{Our Approach} & \textbf{99.21 $\pm$ 0.42} & \textbf{96.43 $\pm$ 0.59} & \textbf{7.68 $\pm$ 0.64} & \textbf{97.58 $\pm$ 0.49} & \textbf{5.85 $\pm$ 0.67} \\ \hline
 & \textbf{GNN\cite{BaselineGNNWiFi_Wang2024}} & 100 & 91.01 & 11.58 & 96.91 & 9.47 \\ \hline
 & \textbf{CFNN\cite{Cha2022}} & 100 & 92.57 & 11.97 & 87.11 & 10.64 \\ \hline
 & \textbf{HADNN\cite{Cha2022}} & 99.46 & 90.01 & 12.01 & 90.72 & 11.12 \\ \hline
 & \textbf{CNNLoc\cite{Song2019}} & 99.82 & 91.56 & 36.68 & 91.78 & 11.50 \\ \hline
\textbf{AP Random Selection} & \textbf{Our Approach} & \textbf{99.12 $\pm$ 0.59} & \textbf{96.26 $\pm$ 0.63} & \textbf{8.42 $\pm$ 0.87} & \textbf{97.08 $\pm$ 0.59} & \textbf{6.74 $\pm$ 0.26} \\ \hline
 & \textbf{GNN\cite{BaselineGNNWiFi_Wang2024}} & 99.82 & 91.01 & 14.07 & 95.13 & 11.23 \\ \hline
 & \textbf{CFNN\cite{Cha2022}} & 99.45 & 90.36 & 13.35 & 83.76 & 10.93 \\ \hline
 & \textbf{HADNN\cite{Cha2022}} & 99.54 & 89.10 & 14.36 & 82.98 & 10.45 \\ \hline
 & \textbf{CNNLoc\cite{Song2019}} & 99.75 & 88.98 & 38.44 & 91.25 & 11.18 \\ \hline
\end{tabular}
\vspace{6pt}
\caption{Comparison Performance of Proposed Work with S.O.T.A on Different AP Selection Methods for 70\% of All APs on UJIIndoorLoc and UTSIndoorLoc Datasets.}
\label{table:compSOTAwithDiffAPSel}
\end{table*}

\subsection{Performance Comparison with Related Works}
 The performance of the proposed framework using the UJIIndoorLoc \cite{UJIIndoorLoc2014} and UTSIndoorLoc \cite{UTSIndoorLoc2023} datasets is evaluated and compared. The focus is improving AP  selection for multi-floor and building environments, with performance assessed using building accuracy, floor accuracy, and mean location error. Tables \ref{table:compSOTA} and \ref{table:compSOTAwithDiffAPSel} compare the proposed method with state-of-the-art (SOTA) approaches for different AP selection strategies.

\subsubsection{Performance on UJIIndoorLoc Dataset} The proposed method achieves 99. 37\% building accuracy, 96. 89\% floor accuracy and a mean location error of 6.29 meters. In comparison, the baseline GNN-based method\cite{BaselineGNNWiFi_Wang2024} achieves perfect building accuracy (100\%) but lags in floor accuracy (92.82\%) and mean location error (9.61 meters). CNNLoc\cite{Song2019} shows significantly lower performance with a mean location error of 35.67 meters, highlighting the limitations of traditional models in multi-floor environments.

\subsubsection{Performance on UTSIndoorLoc Dataset} It is shown that the proposed method continues to excel with 98.12\% floor accuracy and a mean location error of 4.64 meters. The baseline GNN-based method\cite{BaselineGNNWiFi_Wang2024} achieves 100\% building accuracy but performs worse in floor accuracy (97.19\%) and mean location error (7.87 meters). These results underscore the robustness of the proposed approach in diverse indoor environments.

\subsubsection{AP Selection Comparison} Table \ref{table:compSOTAwithDiffAPSel} compares the proposed method with three AP selection strategies: AP Selection Network, AP Threshold Selection, and AP Random Selection. Across all methods, the proposed approach consistently outperforms the SOTA techniques. For example, with AP Threshold Selection in the UJIIndoorLoc dataset, it achieves 99.21\% building accuracy, 96.43\% floor accuracy, and a mean location error of 7.68 meters, outperforming baseline GNN-based method\cite{BaselineGNNWiFi_Wang2024}, which exhibits a higher mean location error (11.58 meters). Similarly, for AP Random Selection, the proposed method achieves 96.26\% floor Accuracy and 8.42 meters mean location error, outperforming baseline GNN-based method\cite{BaselineGNNWiFi_Wang2024} , which shows a mean location error of 14.07 meters and others.

\subsubsection{Discussion} Across both datasets and in all AP selection strategies, the proposed framework demonstrates superior building and floor Accuracy while maintaining consistently low mean location errors. Quantum-assisted data enhancement improves robustness against noise and variability, providing reliable localization even in challenging scenarios. As shown in Fig. \ref{fig:3DVisualResults}, these results establish the proposed method as a robust solution for WiFi-based indoor localization, with strong potential for deployment in real-world multi-floor and multi-building environments.

\begin{figure*}[t]
    \centering
    \includegraphics[width=\textwidth]{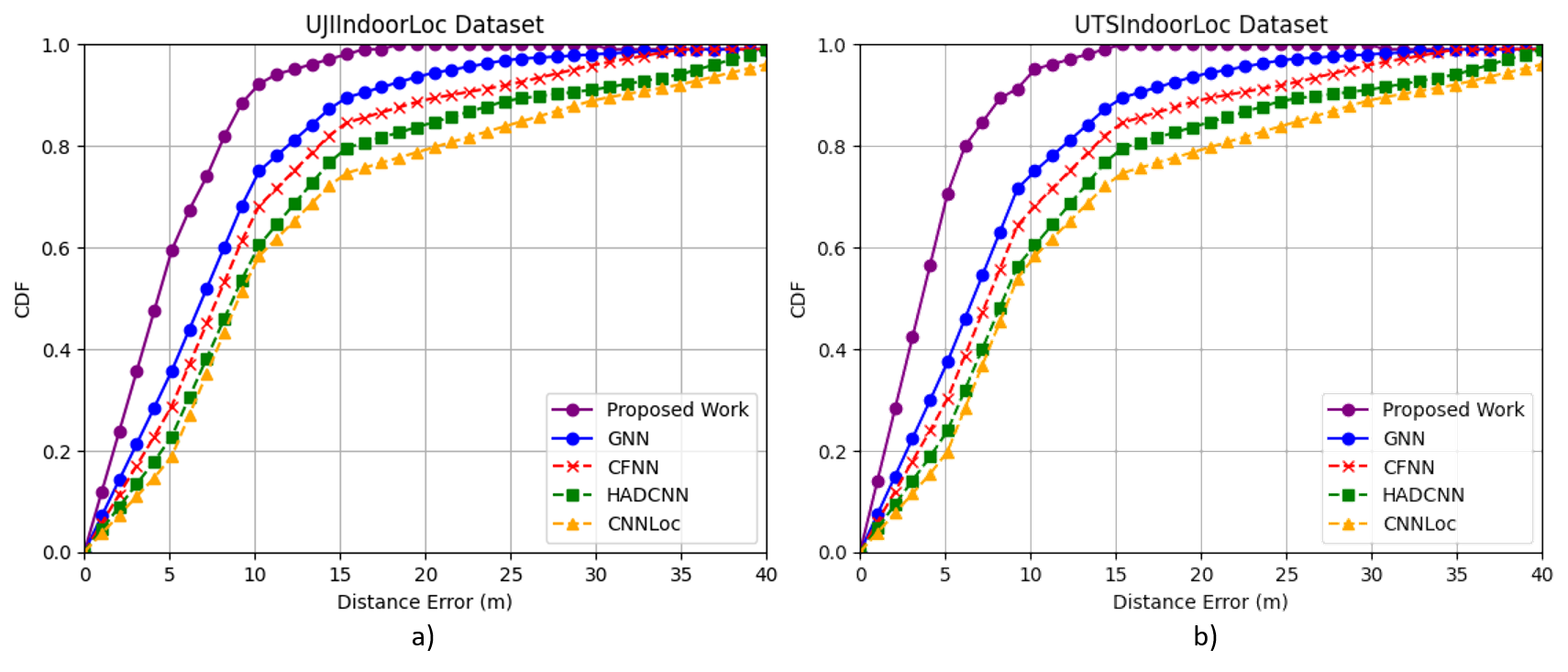} 
    \caption{CDF analysis and comparison of our approach with related works on a) UJIIndoorLoc dataset and b) UTSIndoorLoc dataset.}
    \label{fig:CDFDEBothDatasets}
\end{figure*}

\subsection{CDF Analysis and Comparison}

The cumulative distribution function (CDF) of the distance error for the UJIIndoorLoc and UTSIndoorLoc datasets is illustrated in Fig. \ref{fig:CDFDEBothDatasets}. The analysis focuses on evaluating the effectiveness of the proposed work in minimizing localization errors compared to state-of-the-art models.

\subsubsection{UJIIndoorLoc Dataset} The distance error distribution for the UJIIndoorLoc dataset is shown in Fig. \ref{fig:CDFDEBothDatasets}-a. The proposed work achieves the steepest curve closest to the y-axis, indicating consistently lower distance errors. The baseline GNN-based method\cite{BaselineGNNWiFi_Wang2024} follows with competitive performance, while CFNN and HADCNN\cite{Cha2022} exhibit higher errors, particularly at larger distance ranges. CNNLoc \cite{Song2019} shows the worst performance, with a slower rise in the CDF curve, reflecting significant errors in the dataset.

\subsubsection{UTSIndoorLoc Dataset} Similarly to the UJIIndoorLoc dataset, the proposed work demonstrates the lowest distance error and outperforms all other models (see Fig. \ref{fig:CDFDEBothDatasets}-b). The gap between the proposed work and the baseline GNN-based method\cite{BaselineGNNWiFi_Wang2024} is narrower compared to the UJIIndoorLoc dataset but still favors the proposed work. CFNN and HADCNN\cite{Cha2022} again show higher errors, and CNNLoc \cite{Song2019} exhibits the largest errors even in lower distance ranges.

\subsubsection{Discussion} Across both datasets, the proposed work consistently minimizes distance errors, as evidenced by its superior CDF curves. Although the baseline GNN-based method\cite{BaselineGNNWiFi_Wang2024} shows competitive results, other methods, including CFNN and HADCNN\cite{Cha2022}, and CNNLoc \cite{Song2019}, trail significantly in accuracy. These findings underscore the robustness and effectiveness of the Proposed Work in achieving high-precision indoor localization, particularly in challenging multi-floor environments.

\begin{figure*}[t]
    \centering
    \includegraphics[width=\textwidth]{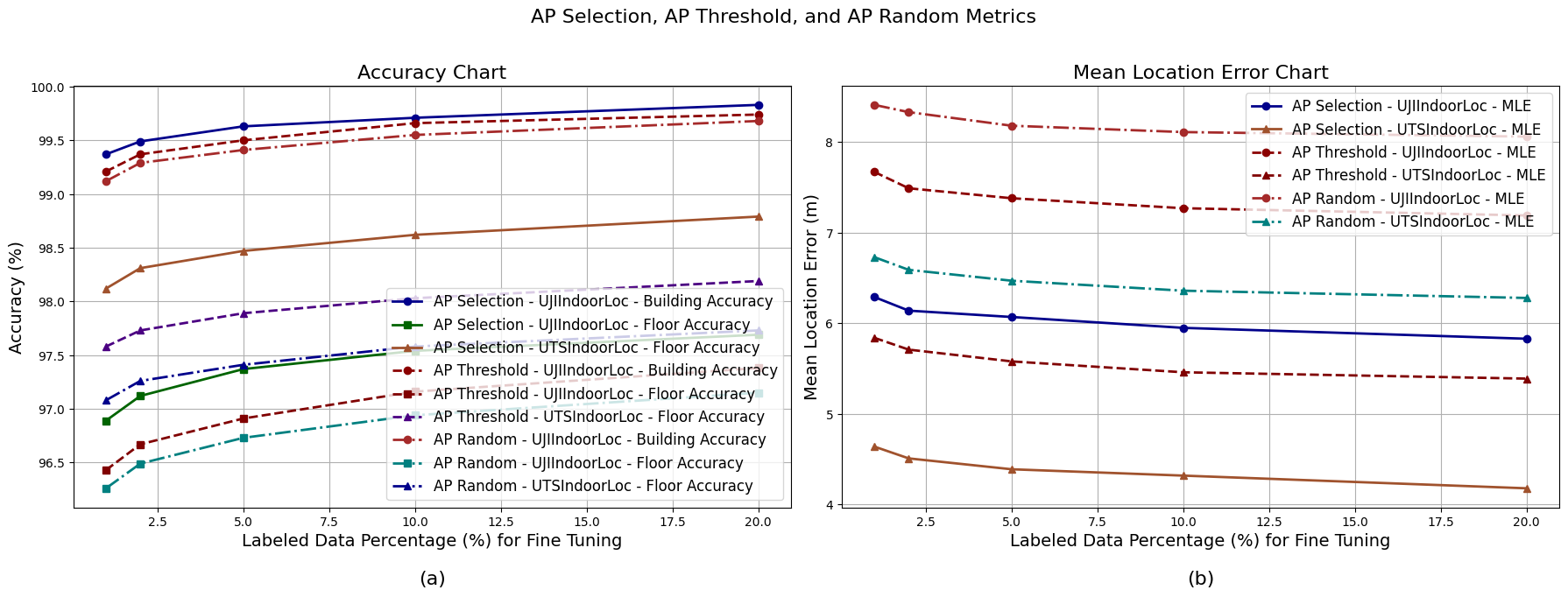} 
    \caption{Impact of Labeled Data with respect to AP Selection, AP Threshold, and AP Random for UJIIndoorLoc and UTSIndoorLoc Datasets.}
    \label{fig:LabeledData}
\end{figure*}

\subsection{Impact of Labeled Data}
The influence of labeled data percentage on mean location error and classification accuracy across different AP selection methods (AP Selection, AP Threshold, and AP Random) for the UJIIndoorLoc and UTSIndoorLoc datasets is evaluated as shown in Fig. \ref{fig:LabeledData}.

\subsubsection{Classification Accuracy} Fig. \ref{fig:LabeledData}-a shows the impact of the percentage of labeled data on the accuracy of classification for floor and building tasks. A positive correlation between labeled data percentage and accuracy is observed, with all methods benefiting from more labeled data. The AP Selection method consistently achieves the highest accuracy across both datasets, reflecting its ability to maximize the utility of labeled data for classification. The AP Threshold method demonstrates steady improvements, particularly for building and floor classification tasks. Meanwhile, while showing improvements with more labeled data, the AP Random method falls behind the other two methods regarding accuracy growth.

\subsubsection{Mean Location Error} The relationship between labeled data percentage and Mean Location Error (in meters) is illustrated in Fig. \ref{fig:LabeledData}-b. As the percentage of labeled data increases, the error consistently decreases across all methods, emphasizing the importance of labeled data for fine-tuning. The AP Selection method achieves the lowest error on both datasets, particularly at higher labeled data percentages, showcasing its efficiency in reducing localization errors. The AP Threshold method also performs well, showing a steady reduction in error, although at a slower rate. In contrast, the AP Random method exhibits higher error rates, indicating that random AP selection is less effective in leveraging labeled data.

\subsubsection{Dataset Comparison} The UJIIndoorLoc dataset slightly outperforms UTSIndoorLoc in building and floor classification accuracy, indicating that it may have more structured or simpler data conducive to better model performance. The AP Selection method for both datasets remains the most effective for minimizing mean location error and maximizing classification accuracy.

\subsubsection{Discussion} These findings highlight the critical role of labeled data in enhancing model performance. The AP Selection method is the most effective strategy, consistently reducing location error and improving classification accuracy. Fine-tuning with labeled data significantly benefits the proposed framework, with clear improvements observed in both metrics as the number of labeled data increases. However, the slower rate of improvement of the AP Random method underscores the value of targeted selection methods like AP Selection and AP Threshold for WiFi-based indoor localization tasks.

\section{Conclusion}
This paper presents Quantum Stochastic Contrastive Learning (QSCL), a framework for WiFi-based indoor localization that addresses data variability, environmental noise, and limited labeled data. Grounded in theory, QSCL uses theorems and a lemma to prove its capabilities in augmentation, diversity, relationship preservation, and quantum noise resilience. By combining quantum-augmented strong data with classical weak augmentations, it enhances diversity and feature learning. An STAA encoder captures spatial and temporal dependencies, while a bidirectional contrastive loss ensures robust representations. Tests on UJIIndoorLoc and UTSIndoorLoc datasets show superior localization accuracy, floor detection, and generalization. QSCL reduces reliance on labeled data and withstands quantum noise, including bit-flip and dephasing. This advances WiFi indoor localization, showing QC and ML's potential in solving real-world challenges. Future work includes implementing QSCL on quantum hardware, expanding scalability, and integrating multimodal data for better accuracy and robustness.

\bibliographystyle{IEEEtran}
\bibliography{references}

\end{document}